\pdfoutput=1
\documentclass[reprint,nofootinbib,aps,prd]{revtex4-2}

\usepackage{graphicx}
\usepackage{dcolumn}
\usepackage{booktabs}
\usepackage{bm}
\usepackage{hyperref}
\usepackage{autobreak}
\usepackage{xcolor}
\usepackage{physics}
\usepackage{amsmath, amssymb} 
\usepackage{mathtools}
\usepackage{ragged2e}
\usepackage{soul} 
\usepackage{orcidlink}

\usepackage{siunitx}
\DeclareSIUnit \parsec {pc}
\newcommand{\ds}{\displaystyle}
\usepackage{tikz,lipsum,lmodern}
\usepackage[T1]{fontenc}
\usepackage[most]{tcolorbox}
\usepackage{lmodern}
\usepackage{wrapfig}

\def\apj{ApJ}                 

\tcbset{colback=green!5,colframe=green!35!black,fonttitle=\bfseries}

\newcommand{\lcdm}{$\Lambda$CDM} 
 
\newcommand{\wowacdm}{$w_0w_a$CDM} 
\newcommand{\wowa}{($w_0, w_a$)} 
\newcommand{\Om}{\Omega_\mathrm{m}}

\newcommand{\Or}{\Omega_\mathrm{R}}

\newcommand{\Obhtwo}{\Omega_\mathrm{b}h^2}
\newcommand{\Ocdmhtwo}{\Omega_\mathrm{cdm}h^2}
\newcommand{\Omhtwo}{\Omega_\mathrm{m}h^2}
\newcommand{\Planck}{\emph{Planck}}

\newcommand{\HnodLCDM}{H_0^{\rm LCDM}}

\newcommand{\DM}{D_\mathrm{M}}
\renewcommand{\DH}{D_\mathrm{H}}
\newcommand{\eV}{\,{\rm eV}}

\newcommand{\kmsMpc}{\,{\rm km\, s^{-1}Mpc^{-1}}}

\begin{document}
\preprint{000-000-000}

\title{Expansion-history preferences of DESI {DR2} and external data}

\author{Prakhar Bansal\,\orcidlink{0009-0000-7309-4341}}
 \email{prakharb@umich.edu}
\author{Dragan Huterer\,\orcidlink{0000-0001-6558-0112}}%
 \email{huterer@umich.edu}
\affiliation{
Department of Physics and Leinweber Center for Theoretical Physics, University of Michigan, 450 Church St, Ann Arbor, MI 48109
}

\date{\today}

\begin{abstract}
We explore the origin of the preference of DESI DR2 baryon acoustic oscillation (BAO) measurements and external data from cosmic microwave background (CMB) and type Ia supernovae (SNIa) that dark energy behavior departs from that expected in the standard cosmological model with vacuum energy (\lcdm). In our analysis, we allow a flexible scaling of the expansion rate with redshift that nevertheless allows reasonably tight constraints on the quantities of interest, and adopt and validate a simple yet accurate compression of the CMB data that allows us to constrain our   phenomenological model of the expansion history.  We find that data consistently show a preference for a {3-4\% increase in the expansion rate at $z\simeq 0.7$}  relative to that predicted by the standard \lcdm\ model, in excellent agreement with results from the less flexible $(w_0, w_a)$ parameterization which was used in previous analyses. Even though our model allows a departure from the best-fit \lcdm\ model at zero redshift, we find no evidence for such a signal.  We also find no evidence (at greater than 1$\sigma$ significance) for a departure of the expansion rate from the \lcdm\ predictions at higher redshifts for any of the data combinations that we consider.  Overall, our results strengthen the robustness  of the  findings using the combination of DESI, CMB, and SNIa data to dark-energy modeling assumptions.
\end{abstract}

\maketitle


\section{Introduction}

Recent constraints on dark energy \cite{DESI:2024mwx, DESI:2025zgx} from Dark Energy Spectroscopic Instrument (DESI \cite{DESI:2016fyo,DESI:2022xcl}) and external data have provided a preference --- though not yet firm evidence --- for dynamical dark energy. Adopting the popular parameterization of the equation of state of dark energy $w(a) = w_0 + w_a(1-a)$ \cite{Linder:2002et,Chevallier:2000qy}, where $w_0$ and $w_a$ are free parameters and $a$ is the scale factor, the analysis of DESI data, combined with cosmic microwave background (CMB) and type Ia supernovae (SNIa) favor values with $w_0 > -1$ and $w_a < 0$, and depart from the standard cosmological model with vacuum energy ($w_0=-1, w_a=0$) at the statistical level between {2.7$\sigma$ and 4.2$\sigma$ \cite{DESI:2025zgx}.} 

One interesting feature implied by the combined analysis of DESI and external data in the \wowacdm\ model \cite{DESI:2024mwx} and some alternative parameterizations \cite{DESI:2024kob,DESI:2024aqx} is that dark energy starts out ``phantom'' (with $w(z)<-1$) at high redshift, and cross into the $w(z)>-1$ regime at $z\lesssim 1$. This at face value implies \textit{two} anomalies not expected in the standard \lcdm\ model: 1) a dark energy density that increases in time (in the phantom regime at higher redshifts), but also  2) dark energy density that subsequently \textit{decreases} in time (in the $w(z)>-1$ regime). Much has been written about these results, with the emphasis on alternative parameterizations of dark energy sector that allow more degrees of freedom \cite{Mukherjee:2024ryz,Dinda:2024ktd,Reboucas:2024smm}, explorations of modified-gravity fits to the data \cite{Ishak:2024jhs,Chudaykin:2024gol}, relation to neutrino-mass constraints \cite{Green:2024xbb, Elbers:2024sha}
and other investigations \cite{Cortes:2024lgw,Berghaus:2024kra,Chan-GyungPark:2024mlx,Poulin:2024ken,Jiang:2024xnu,Wolf:2024eph,Chan-GyungPark:2024brx,Linder:2024rdj,Lewis:2024cqj}. These studies generally confirmed the aforementioned physical picture obtained in the simple $(w_0, w_a)$ parameterization. 

A key question here is whether there is really separate evidence for either $w(z)>-1$ at lower redshifts or for $w(z)<-1$ at higher redshifts --- or for both. Unfortunately most of the studies carried out thus are not equipped to answer this question, as their rigid parameterizations impose a coherence across redshift and may not have enough probing power to detect statistically significant preferences in the data. For instance, in the $(w_0, w_a)$ model, a preference by data for $w(z)>-1$ at low redshift, combined with the general preference to return to the \lcdm\ value ($w(z)\simeq -1$) at $z\simeq 1$ in order to fit e.g.\ the distance to recombination, \textit{guarantees} a preference for phantom dark energy at $z\gtrsim 1$ just because of the stiffness of the parameterization. Richer descriptions of the dark-energy sector, including a binned description of the equation of state, are possible, but suffer from large parameter-space degeneracies, and consequently poor constraints. Another, very different, approach discussed in the community was removing individual data-points (in e.g.\ DESI's measured distances) and seeing how the constraints change, but this is a very inefficient way to understand what the data are really telling us, and specifically \textit{where} the preference for dynamical energy is  coming from. 

Our goal here is to identify precisely what features in the data are responsible for the preference for dynamical dark energy seen by DESI and external data. To that effect, we choose to consider a piecewise-constant parameterization of the expansion rate $H(z)$. This approach has several  advantages: 1) like the equation of state $w(z)$, it too directly propagates into the different kinds of distances 
probed by DESI data, and communicates the effects of dark energy to SNIa and CMB observables (see the next section for details); 2) unlike any smooth equation-of-state description, direct parameterization of $H(z)$ allows rapid changes in the expansion rate, which is especially useful in understanding preferences shown by the data at low redshift, and 3) despite its flexibility, a binned $H(z)$ prescription does not suffer from large degeneracies and allows us to constrain all model parameters, and the derived distances, to a reasonably good precision. We are reluctant to follow a tiresome and inaccurate tradition and call this method ``model-independent'', but the built-in variation of the expansion rate in multiple redshift bins is the key feature that allows us to understand how different redshift ranges contribute to the results.

\section{Model and analysis methodology}

We consider a model that modifies the standard Hubble parameter by incorporating perturbative parameters, allowing us to explore deviations from the standard $\Lambda$CDM
framework across multiple redshift bins. We refer to this model as the modified-H model; it is given by
\begin{equation}
    H(z) = \HnodLCDM E(z)\,(1+\alpha_i),
    \label{eq:Hz}
\end{equation}
where $\HnodLCDM$ is the Hubble constant in the standard \lcdm\ model and $\alpha_i$ are the parameters of the model whose nonzero values allow departures from \lcdm. We define each $\alpha_i$ to be defined in one of the six redshift bins which  approximately\footnote{Our binning is identical to that from DESI DR2 \cite{DESI:2025zgx}, except that we extend the lowest bin, which was originally $0.1<z<0.4$, all the way down to $z=0$ in order to allow the Hubble parameter at $z=0$ to vary independently of $\HnodLCDM$, as well as to take into account the impact of SNIa data at $z<0.1$.} coincide with the six DESI DR2 bins into which the BAO measurements were compressed \cite{DESI:2024mwx} (see their Table \ref{tab:params}). The function $E(z)$ is given in its standard \lcdm\ form
\begin{align}
E(z) &= \sqrt{\Om(1+z)^3+\Or(1+z)^4+(1-\Om-\Or)}\notag\\[0.2cm]
\Om &\equiv \ds\left (\frac{100}{\HnodLCDM}\right )^2 \left (\Ocdmhtwo+\Obhtwo\right )\,,
    \label{eq:Ez}
\end{align}
where the Hubble constant is in units of $\kmsMpc$, $\Ocdmhtwo$ and $\Obhtwo$ are the physical energy densities in cold dark matter and baryons respectively. Appendix \ref{app:mnu} explains how massive neutrinos contribute to radiation density $\Or$, scaling as nonrelativistic matter at low redshifts and becoming progressively more relativistic at high $z$.  In Eqs.~(\ref{eq:Hz}) and (\ref{eq:Ez}), we set $\HnodLCDM$ to its best-fit alue from the standard \lcdm\ analysis with the alpha parameters set to zero, which is about  $68\kmsMpc$, the precise value depending on the data combination used (see the next Section). Moreover, note from Eq.~(\ref{eq:Ez}) that $\Om$ is a derived parameter in our analysis. 
The transverse and line-of-slight comoving distances are respectively
\begin{equation}
    \DM(z) = c\int_{0}^{z}  \frac{dz^\prime}    {H(z^\prime)}; 
    \quad 
    \DH(z) = \frac{c}{H(z)}~.
    \label{eq:DM_DH}
\end{equation}

 We assume that the model reverts to standard \lcdm\ at redshifts above the highest bin, at $z>4.16$. No such assumption has been made at low redshift, where $H(z=0)$ is allowed to differ from the \lcdm\ value $\HnodLCDM$ even at $z=0$. In particular, the Hubble constant that enters the distances is $H(z=0)= \HnodLCDM (1+\alpha_1)$, which is in general different from the expansion rate that converts from the physical density $\Omhtwo$ to matter density relative to critical $\Om$ (see again Eq.~\ref{eq:Ez}). This agrees with our logic that both $\Omhtwo$ and $\Om$ are set in the early universe where the unmodified Hubble parameter $\HnodLCDM E(z)$ is relevant.

With all that in mind, there are a total of eight parameters in our analysis: the six alphas, and the physical baryon and CDM densities $\Obhtwo$ and $\Ocdmhtwo$ which control the sound horizon as well as enter the lower-redshift distances as in Eq.~(\ref{eq:Hz}). These parameters and their respective priors in our analysis are given in Table \ref{tab:params}.

\renewcommand{\arraystretch}{1.2}
\begin{table}[t]
\caption{\label{tab:params}Parameters used in our modified-H analysis and their respective (flat) prior ranges. Note that $\HnodLCDM$ is only varied in our \lcdm\ analysis (which we run for comparison and where we also vary $\Obhtwo$ and $\Ocdmhtwo$); it is fixed to the \lcdm's best value in the modified-H analysis. See text for details. {The redshift bins for $\alpha$ parameters are defined in \textit{left-closed, right-open} intervals}  }
\begin{ruledtabular}
\begin{tabular}{lcc}
\textrm{Parameter}&
\textrm{Description}&
\textrm{Prior (flat)}\\
\colrule
$\Obhtwo$ & Physical baryon density & [0.005,0.1]\\
$\Ocdmhtwo$ & Physical CDM density & [0.002,0.20]\\
$\HnodLCDM$& unmodified Hubble constant& [20,100]\\
$\alpha_1$& $0.0\le z<0.4$ & $[-0.1,0.3]$\\
$\alpha_2$& $0.4\le z<0.6$ & $[-0.1,0.15]$\\
$\alpha_3$& $0.6\le z<0.8$ & $[-0.1,0.15]$\\
$\alpha_4$& $0.8\le z<1.1$ & $[-0.1,0.1]$\\
$\alpha_5$& $1.1\le z<1.6$& $[-0.1,0.1]$\\
$\alpha_6$& $1.6\le z<4.16$ & $[-0.1,0.1]$\\
\end{tabular}
\end{ruledtabular}
\end{table}

We modify the standard cosmological code \texttt{CAMB} to compute the background quantities like Hubble rate, distances and supernova magnitudes in our modified-H model. To obtain the constraints on the cosmological parameters of our model we use Monte Carlo Markov Chain (MCMC) sampler in \texttt{Cobaya} \cite{Torrado:2020dgo}. For our MCMC chains, we use the default convergence criteria of Gelman and Rubin R statistic $<$  0.01. To calculate the means, confidence intervals and likelihood distributions for our model parameters, we use \texttt{GetDist} \cite{Lewis:2019xzd} code with our converged MCMC chains.

\section{Data}

We use the following data:\\

\textbf{DESI DR2 BAO.}
{We use the measurements from the DESI Data Release 2 BAO analysis (henceforth DESI DR2 BAO, or just DESI), and adopt the 13 distance measurements, and their covariance, as quoted in Table IV of \cite{DESI:2025zgx} and validated in supporting DESI DR2 publications \cite{DESI:2025qqy,DESI:2025qqu}.} To make our analysis simple and as model-independent as possible, we do not use the additional information from the full-shape clustering of DESI sources \cite{DESI:2024hhd}.\\

\textbf{Compressed CMB data.} It is often very useful to compress the CMB data to a few physically motivated quantities. There are two fundamental reasons for this: first, such compression allows analyses of purely phenomenological models for which a theoretically expected CMB angular power spectrum cannot be computed. And second, the compression also allows a much faster evaluation of the CMB likelihood than a full power-spectrum-based likelihood would. {The compression is likely to accurately capture information from dark-energy models that smoothly affect the expansion and growth history.}

Following a similar well-established approach (e.g.\ \cite{Bond:1997wr,Wang:2007mza, Elgaroy:2007bv,WMAP:2010qai,Vonlanthen:2010cd,Huang:2015vpa,Zhai:2019nad}), we compress the CMB into three physical quantities: the ``shift'' parameter $R$ \cite{Bond:1997wr} and the angular location $\ell_a$, which are defined as
\begin{equation}
  \begin{aligned}
    R &= 100\sqrt{\Obhtwo+\Ocdmhtwo+\Omega_{\nu,\mathrm{m}} h^2}D_{M*}/c\\[0.1cm]
    \ell_a &= \pi D_{M*}/r_*,
    \end{aligned}
    \label{eq:R_la}
\end{equation}  
as well as the physical baryon density $\Obhtwo$. Here $D_{M*}$ and $r_*$ are respectively the transverse comoving distance to, and the sound horizon at, the surface of last scattering evaluated at $z_*=1090$. Moreover, ${\Omega_{\nu,\mathrm{m}} h^2}(=\sum m_\nu/93.14)$ describes the massive neutrino density; in this work, we have fixed $\sum m_\nu$ to $0.06 \eV$.

We use the combined likelihood from \Planck\ and Atacama Cosmology Telescope (ACT). Specifically, we adopt the joint likelihood that makes use of the PR3 \Planck\ \texttt{plik} likelihood \cite{Planck_PR3} and the Data Release 6 of ACT \cite{ACT:2023kun}.\footnote{The likelihood is available from \url{https://github.com/ACTCollaboration/act_dr6_lenslike}.}. The resulting compressed datavector is 
\begin{equation}
\mathbf{v}_\mathrm{CMB} \equiv 
\begin{pmatrix}
R \\
\ell_a \\
\Obhtwo 
\end{pmatrix}
=
\begin{pmatrix}

1.7504\\
301.77\\
0.022371\\
\end{pmatrix}
\label{eq:cmb_vec}
\end{equation}\\
and the covariance matrix between these three compressed parameters is given by
\begin{equation}
\begin{aligned}
\mathcal{C}_\mathrm{CMB} = 10^{-8} \times
\begin{pmatrix}
1559.83 & -1325.41 & -36.45  \\
-1325.41 & 714691.80 & 269.77  \\
-36.45 & 269.77 & 2.10 \\
\end{pmatrix}\,.
\end{aligned}
\label{eq:covariance_matrix}
\end{equation}
We find an excellent fit not only for the \lcdm\ model on which this compression was derived, but also for  the \wowacdm\  model. We provide the details of this validation in Appendix \ref{app:cmb}. \\ 

\textbf{Type Ia supernovae.}
Our principal SNIa dataset is the Dark Energy Survey Year 5 Data Release (DESY5 \cite{DES:2024jxu}). It contains 1829 SNIa, of which 1635 are photometrically-classified objects in the redshift range $0.1<z<1.3$, complemented with 194 low-redshift SNIa in the range $0.025<z<0.1$. We also consider two other SNIa datasets (following the same logic in \cite{DESI:2024mwx}): the Union3 compilation of 2087 SNIa \cite{Rubin:2023ovl}, and the PantheonPlus compilation of 1550 spectroscopically-confirmed SNIa in the redshift range $0.001<z<2.26$ \cite{Scolnic:2021amr},  many (1363) in common with Union3. In the PantheonPlus analysis, we also impose a $z>0.01$ condition to object selection in order to mitigate the impact of peculiar velocities in the Hubble diagram \cite{Peterson:2021hel}. In all SNIa data combinations, we marginalize analytically over the offset in the Hubble diagram $\mathcal{M}$ which is a nuisance parameter in a cosmological SNIa analysis.

\begin{figure}[t]
    \centering
    \includegraphics[width=\linewidth]{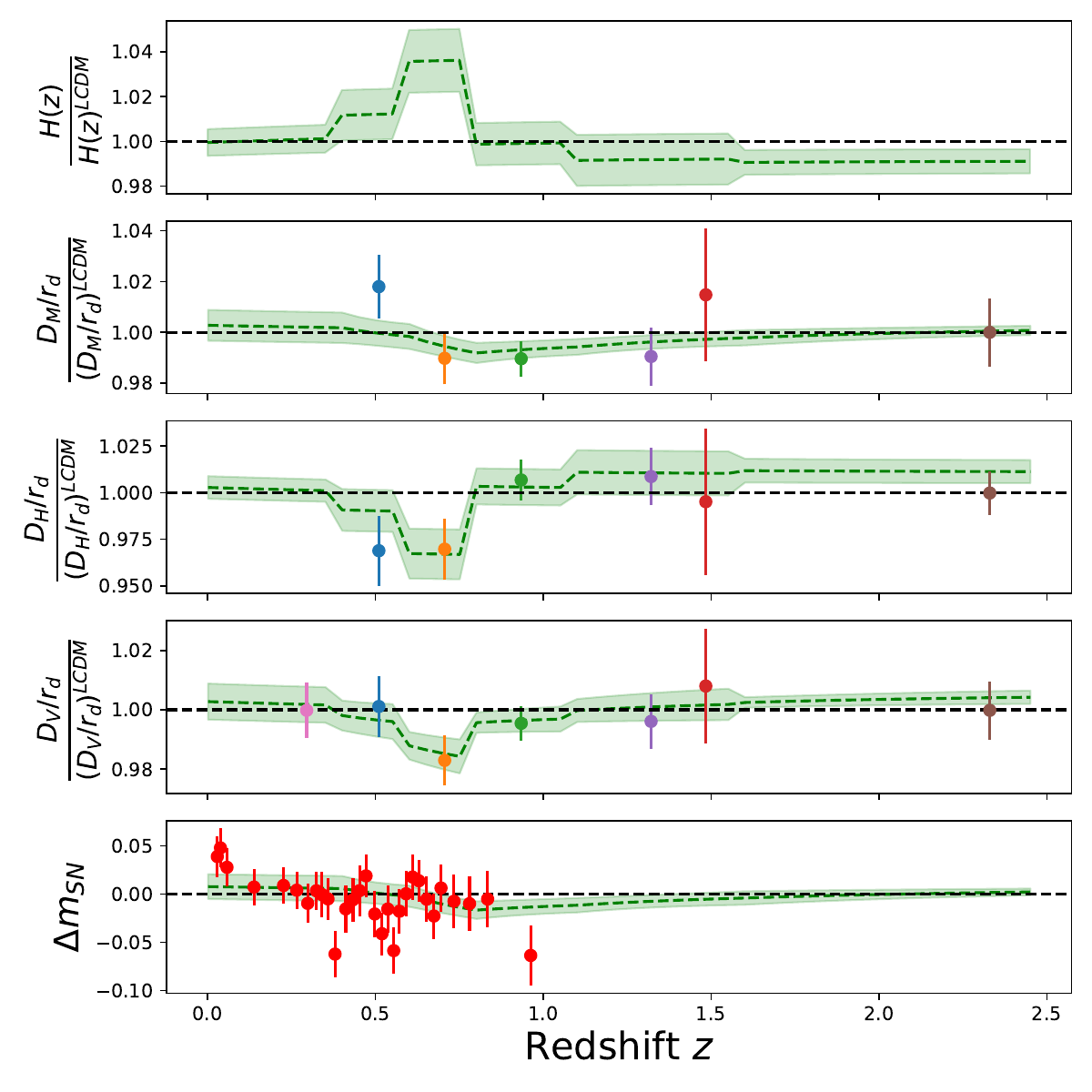} 
    \caption{Constraints from our modified-H model, assuming DESI+CMB+DESY5 data. We show the constraints on the expansion rate $H(z)$ (top panel), followed (in panels that follow, moving down) by the derived constraints on the angular-diameter distance, Hubble distance, and volume-averaged distance all divided by the sound horizon, and finally the apparent magnitude of SNIa. All of the quantities are shown relative to their \lcdm\ values computed with best-fit parameters from our analysis. The data points show the DESI DR2 BAO measurements, except in the lowest panel where we show data from SNIa. See text for more details, and in particular the explanation of how SNIa magnitude residuals were defined.
    }
    \label{fig:Hz_Dz_dm}
\end{figure}

\section{Results}

The constraints on the six alpha parameters from the DESI+CMB+DESY5 analysis, marginalized over the baryon and CDM number densities and the Hubble constant, are shown in Fig.~\ref{fig:alphas} in Appendix \ref{app:alphas}, while the corresponding parameter constraints are shown in Table \ref{tab:constraints}. {The same Appendix also contains more information about our model fitting. }

Overall, we find that the modified-H model gives a slightly better fit to the data than the $(w_0, w_a)$ model --- not nearly better enough to indicate a \textit{preference}  for our more complex model, but sufficiently so to indicate that the model is a good fit to the data. Overall, the alpha parameters are consistent with zero values predicted by \lcdm, though we observe a modest 2$\sigma$ deviation from zero in the {third} alpha parameter. 

A more detailed picture can be obtained by looking at the derived constraints on the Hubble parameter and distances, shown in Fig.~\ref{fig:Hz_Dz_dm} for the DESI+CMB+DESY5 data combination. We show, from top to bottom, the derived Hubble parameter; the corresponding constraints on the angular-diameter, Hubble, and volume-averaged distance; and finally the constraint on the apparent magnitude of SNIa, all as a function of redshift. In all cases, we show constraints relative to their \lcdm\ values computed with best-fit parameters from our analysis (i.e.\ best-fit $\HnodLCDM$ and $\Om$, effectively); the \lcdm\ best-fit values are shown as black dashed lines and centered respectively at either 1 or zero\footnote{There is some ambiguity as to how to show SNIa data compared to two theory models (\lcdm\ and our modified-H model), given that the Hubble-diagram residuals shown here depend on the Hubble-diagram offset $\mathcal{M}$ which, however, has already been marginalized over in the analysis. In the bottom panel of Fig.~\ref{fig:Hz_Dz_dm}, we choose to show the residuals of the best-fit modified-H model  relative to best-fit \lcdm\ by including the respective values of $\mathcal{M}$ which we compute by fitting data \textit{after} the best-fit cosmological parameters in the combined DESI+CMB+SNIa analysis have been determined (and fixed). Similarly, we show the data after subtracting the best-fit \lcdm\ magnitude and the corresponding $\mathcal{M}$.} Note the general agreement of the predictions from our modified-H model with those from \lcdm. The most noticeable discrepancy between the two is a {3-4\% ``bump'' in the expansion rate at $z\simeq 0.7$}, which integrates to contribute to a trough in the distances as {$z\gtrsim 1.6$}.

\begin{figure}[t]
    \centering
    \includegraphics[width=\linewidth]{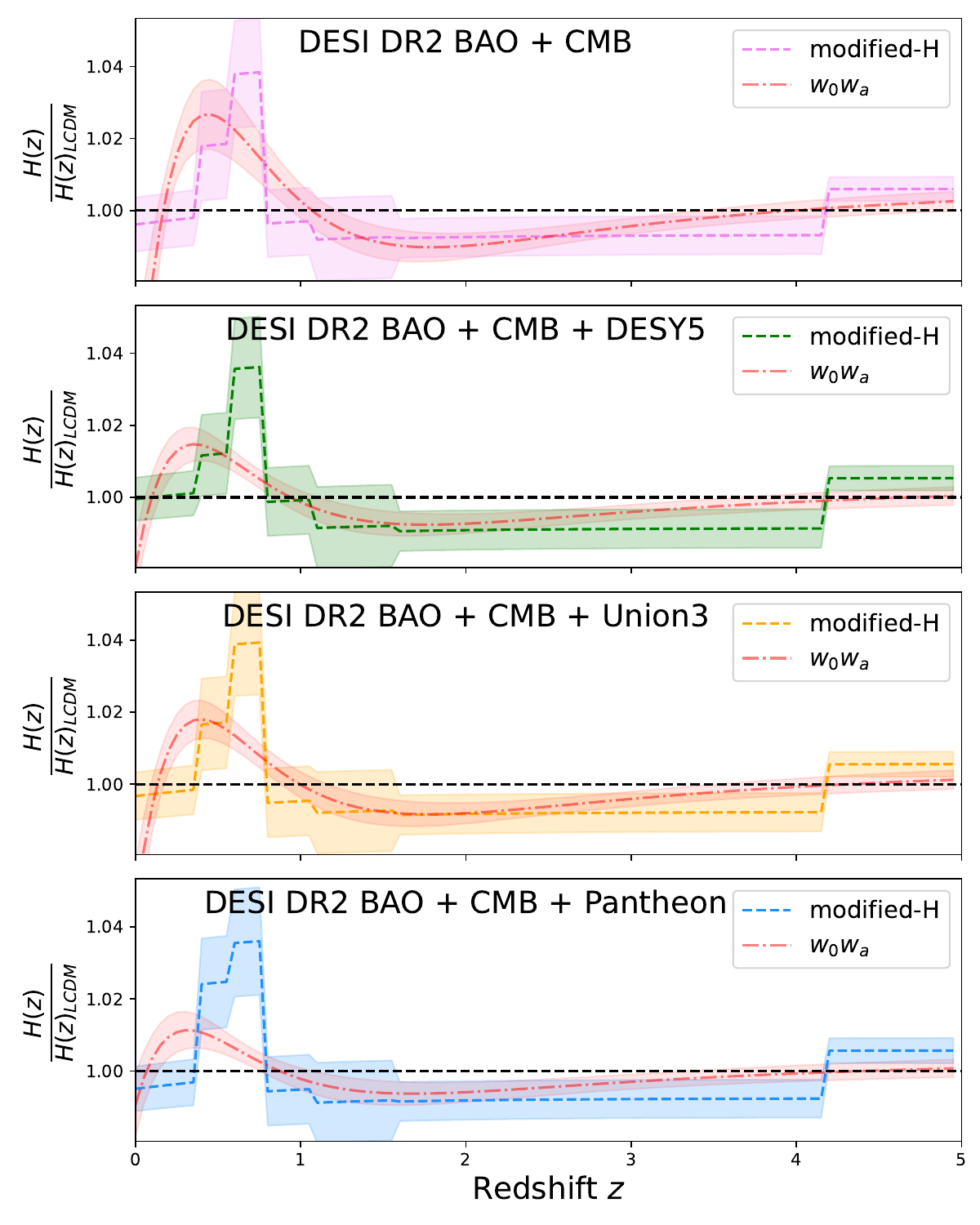}
    \caption{A more detailed view of the $H(z)$ constraint relative to its best-fit value in \lcdm. We show constraints from DESI BAO and CMB in the first panel (without supernovae), followed by the constraints including one of the three SNIa datasets (DESY5, Union3, and PantheonPlus, respectively in the second, third, and fourth panels). In each case we show  comparison with the corresponding best-fit constraint that assumes the  \wowa\ model. 
    }
    \label{fig:H_and_rho}
\end{figure}

Figure \ref{fig:H_and_rho} focuses attention on the preferences in the expansion rate provided by the DESI+CMB and DESI+CMB+SNIa data in our modified-H model. Here, we present the expansion rate relative to that of the best-fit \lcdm\ model for the combination of DESI DR2 BAO and CMB alone, and also DESI and CMB combined with either one of the three SNIa datasets: DESY5, Union3, and PantheonPlus (so the second panel of Fig.~\ref{fig:H_and_rho} has the same information as the top panel in Fig.~\ref{fig:Hz_Dz_dm}). We compare our results with those from the \wowa\ parameterization, shown as the red contours in Fig.~\ref{fig:H_and_rho}. The preferences from the modified-H model show excellent overall agreement with those from the \wowa\ parameterization, both for the DESI BAO+CMB combination and for the combinations that include supernovae datasets (DESI BAO+CMB+SNIa): both models show preference for the aforementioned $\sim$3-5\% bump in the expansion rate at {$z\simeq 0.5 - 0.7$ }. This bump is not particularly significant (being {2.6$\sigma$ for the DESI BAO + CMB combination, and 2.6$\sigma$, 2.7$\sigma$, and 2.4$\sigma$ respectively for combinations of DESI and CMB with DESY5, Union3, and PantheonPlus  }), but it importantly agrees with the same feature in \wowa\ model which does not have the flexibility to cleanly isolate this feature but does have more statistical power due to having fewer parameters. Further, both the modified-H and \wowa\ model show a very mild ($\sim$1$\sigma$)  preference for a lower-than-Lambda $H(z)$ at $z\gtrsim 1.5$. 

The late-time increase in the expansion rate, if confirmed by future data, would correspond to the corresponding increase in the dark-energy density. Roughly speaking, such an increase could be caused by ``thawing'' scalar field \cite{Caldwell:2005tm} that starts to evolve at late times ($z\lesssim 1$) and thus has an equation of state $w(z)>-1$ and a correspondingly higher density than that in vacuum energy. We do not pursue comparisons with specific dark-energy models further. 

One particularly interesting and, to our knowledge, novel result is that all four data combinations shown in Fig.~\ref{fig:H_and_rho} favor the model where $H(z)$ agrees with the \lcdm\ prediction (anchored at high redshift) even at very low redshift, as $z\rightarrow 0$. The \wowa\ parameterization does not allow such a variation, since $H(z)$ is affected by a  change in $w$ only at first order in $z$ (and comoving distance at second order) \cite{Huterer:2000mj}; in other words, the value of $H(z=0)$ is completely determined by values of $\HnodLCDM$ and $\Om$ anchored at high redshift in both \lcdm\ and \wowacdm\ model. In contrast, our parameterization allows a more abrupt, zeroth-order in redshift change in $H(z)$. This in principle allows the disagreement between $H(z\rightarrow 0)$ and $\HnodLCDM$ in the modified-H model, which the data however do not prefer. It is tantalizing to consider implications of this internal concordance test for direct Hubble-constant measurements that use the astronomical distance ladder, but that will require an in-depth investigation that we leave for near-future work.

Finally, we note that we have done internal checks by changing the details of the binning. We found some dependence of the $\Delta\chi^2$ values on the choice of the binning, but overall results that are consistent with those presented here.

\section{Conclusions}

Our principal goal, with the flexible modeling of the expansion history and dark-energy sector given in our Eq.~(\ref{eq:Hz}), was to give a somewhat more nuanced conclusions on dark energy than constraints from simple parameterizations of the equation of state of dark energy. Assuming the combination  of DESI DR2 BAO, compressed \Planck\ and ACT, and SNIa data, we find results remarkably consistent with those from the popular \wowa\ model. We observe a mild preference for the $\sim$3-4\% ``bump'' in the expansion rate at $z\simeq 0.7$ relative to the fit of \lcdm\ model to the same data, and a general agreement with findings from \lcdm\ at higher redshift. Moreover, even though we allow variations in the expansion rate at low redshift  relative to $H(z)$ anchored at high redshift, we see no evidence for the departure from \lcdm\ model's expectation as $z\rightarrow 0$. 

One might be tempted to criticize our modified-H model as insufficiently ``physical'', as it allows for sharp transitions in the expansion rate. We think that our model's flexibility, especially one that goes beyond that of smooth $w(z)$ descriptions, is precisely its feature. Given the lack of \textit{any} compelling dark-energy models,  it is  essential to keep an open mind regarding the description of the dark-energy sector. That is what we have proceeded to do here. 

Finally, we have also provided in Appendix \ref{app:cmb} an accurate  compression of the CMB (\Planck+ACT) data, which should prove useful in constraining beyond-standard models of dark energy.

It is becoming clear that a relatively low redshift range, $z\lesssim 0.8$, is becoming very interesting to explore with future and better data. This is where we (with the modified-H model) and previous findings (with the \wowa\ parameterization) consistently see hints of a preference for dynamical dark energy, although this could certainly be just a statistical fluctuation. This is also the redshift range where we rely in good measure to data from SNIa, which are essential in predicting the behavior of dark energy at $z\lesssim 1$. It is precisely at these relatively low redshifts where additional BAO data will be extremely useful. 

\vspace{0.5cm}
\textit{Acknowledgments.}
This work has been supported by the Department of Energy under contract DE‐SC0019193. We thank our many collaborators on the DESI analysis team for discussions that have influenced our thinking about the implications of DESI dark-energy results.

\appendix
\section{Validation of the CMB compression}\label{app:cmb}

Here we provide more details on the validation of our compressed CMB datavector and its covariance shown in Eqs.~(\ref{eq:cmb_vec})-(\ref{eq:covariance_matrix}).

We obtain the aforementioned compression by considering the joint likelihood that makes use of the PR3 \Planck\ \texttt{plik} likelihood \cite{Planck_PR3} likelihood \cite{Carron:2022eyg} and the Data Release 6 of ACT \cite{ACT:2023kun}.
We next run the Monte Carlo Markov Chain (MCMC) sampler  \texttt{Cobaya} \cite{Torrado:2020dgo} on these data. We compute the three-dimensional compressed data-vector and its covariance by running \texttt{Cobaya} on the \lcdm\ model. Then, for our validation tests, we use the actual \lcdm\ constraints from the aforementioned analysis, and also run \texttt{Cobaya} on the \wowacdm\ model to get constraints in that parameter space.

\begin{figure*}[t]
    \centering
    \includegraphics[width=0.49\linewidth]{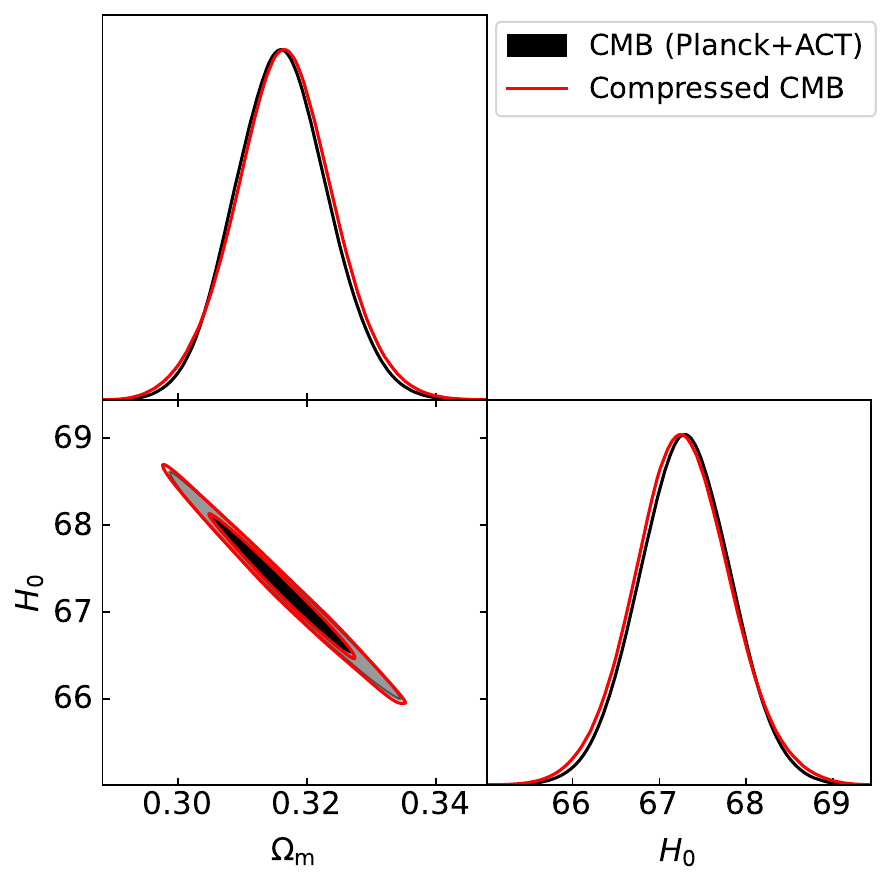}
    \includegraphics[width=0.49\linewidth]{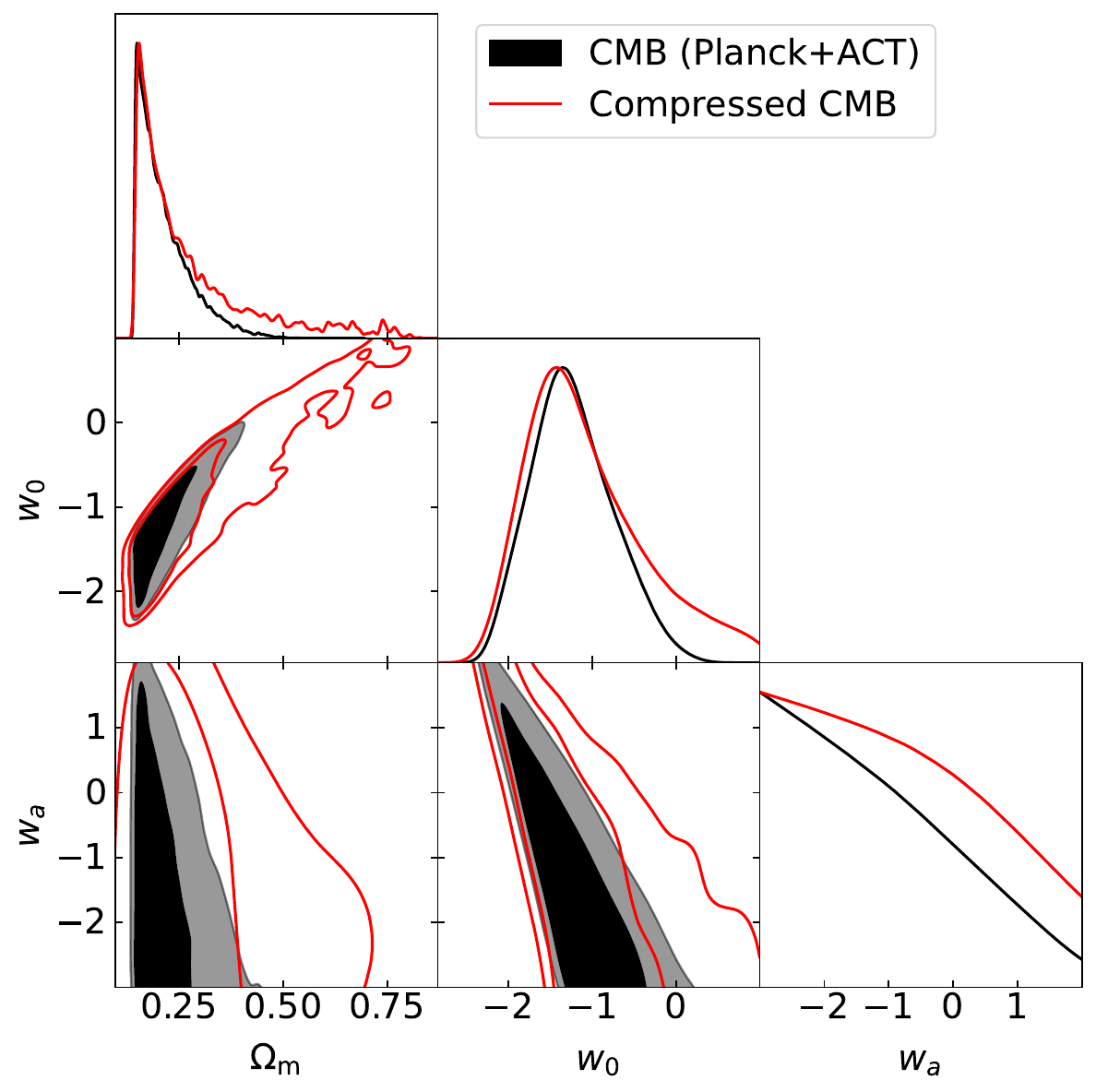}    
    \caption{Comparison of full CMB constraints  to those from the compressed CMB datavector in the \lcdm\ model (left panel) and \wowacdm\ model (right  panel). Black lines and filled contours show the constraint from the full \Planck\ (2018) + ACT (DR6) chain, while the red lines and open contours show the approximate constraints from our compressed data-vector. {Note that the constraints in the right panel are poor because we use CMB data alone in this Figure, rather than in combination with the BAO.}}
    \label{fig:cmb_constraints}
\end{figure*}

For our MCMC chains, we use the default convergence criteria of Gelman and Rubin R statistic $<$  0.01. To calculate the means, confidence intervals and likelihood distributions for our model parameters, we use \texttt{GetDist} \cite{Lewis:2019xzd} code with our converged MCMC chains.

Figure \ref{fig:cmb_constraints} shows the comparison of the constraints from the CMB and those derived from our compressed datavector in two cosmological models. In the left panel, we show constraints in \lcdm\, with the matter density relative to critical $\Om$ and the Hubble constant $H_0$ as the only free parameters. The agreement between the full \Planck\ (2018)+ACT (DR6) likelihood is excellent, though mostly by construction since our compressed quantities are derived from the \lcdm\ MCMC analysis.

A much more convincing validation is obtained with a comparison to another model on which the datavector was not explicitly trained.  We adopt the popular phenomenological model that models the equation of state of dark energy as $w(a) = w_0+w_a(1-a)$, where $a$ is the scale factor and $w_0$ and $w_a$ are two free parameters. In the right panel of Fig.~\ref{fig:cmb_constraints} we compare constraints on the parameter space $(\Om, w_0, w_a)$ from the full CMB chain and those derived with our compressed datavector. The agreement is now visually less good, especially in $w_a$, but note that the principal features of the CMB constraint in this very degenerate parameter space are still reasonably well recovered with the compressed analysis. This bodes well for more realistic cosmological analyses when various datasets will be combined with the CMB in parameter spaces that would be very poorly determined by the CMB alone. 

\begin{figure*}[t]
    \centering
    \includegraphics[width=0.49\linewidth]{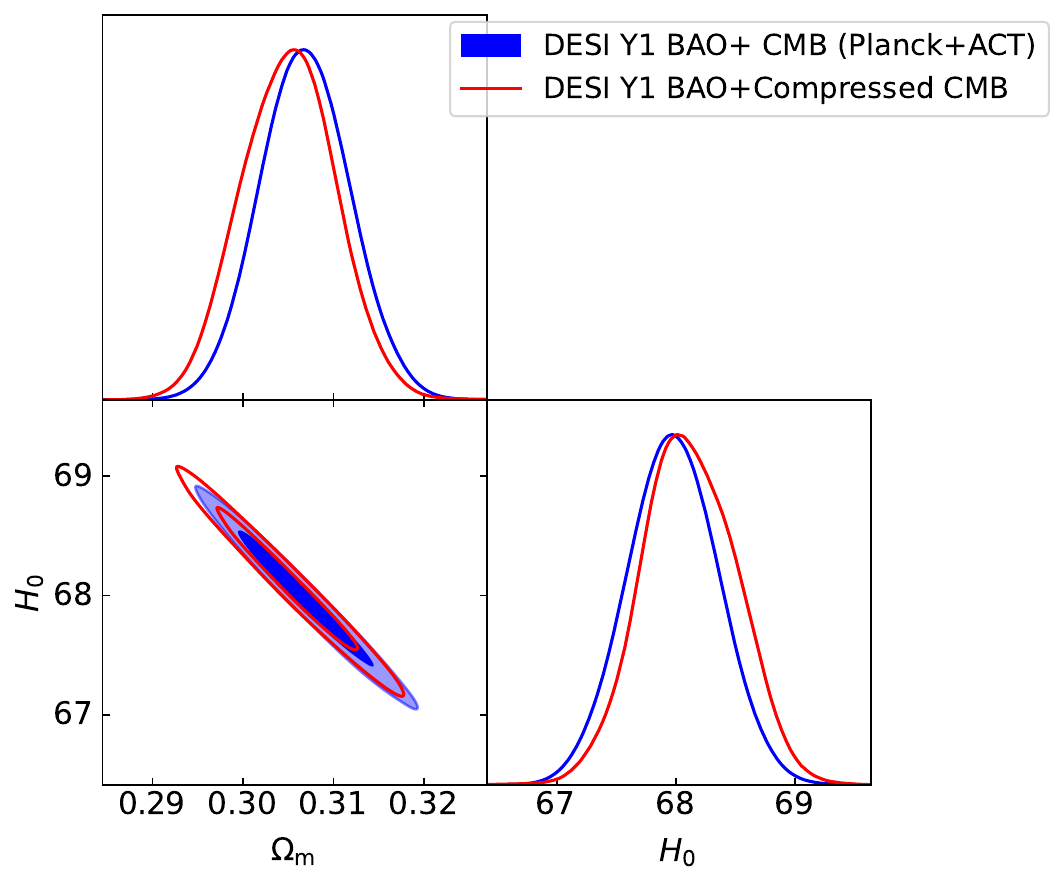}
    \includegraphics[width=0.49\linewidth]{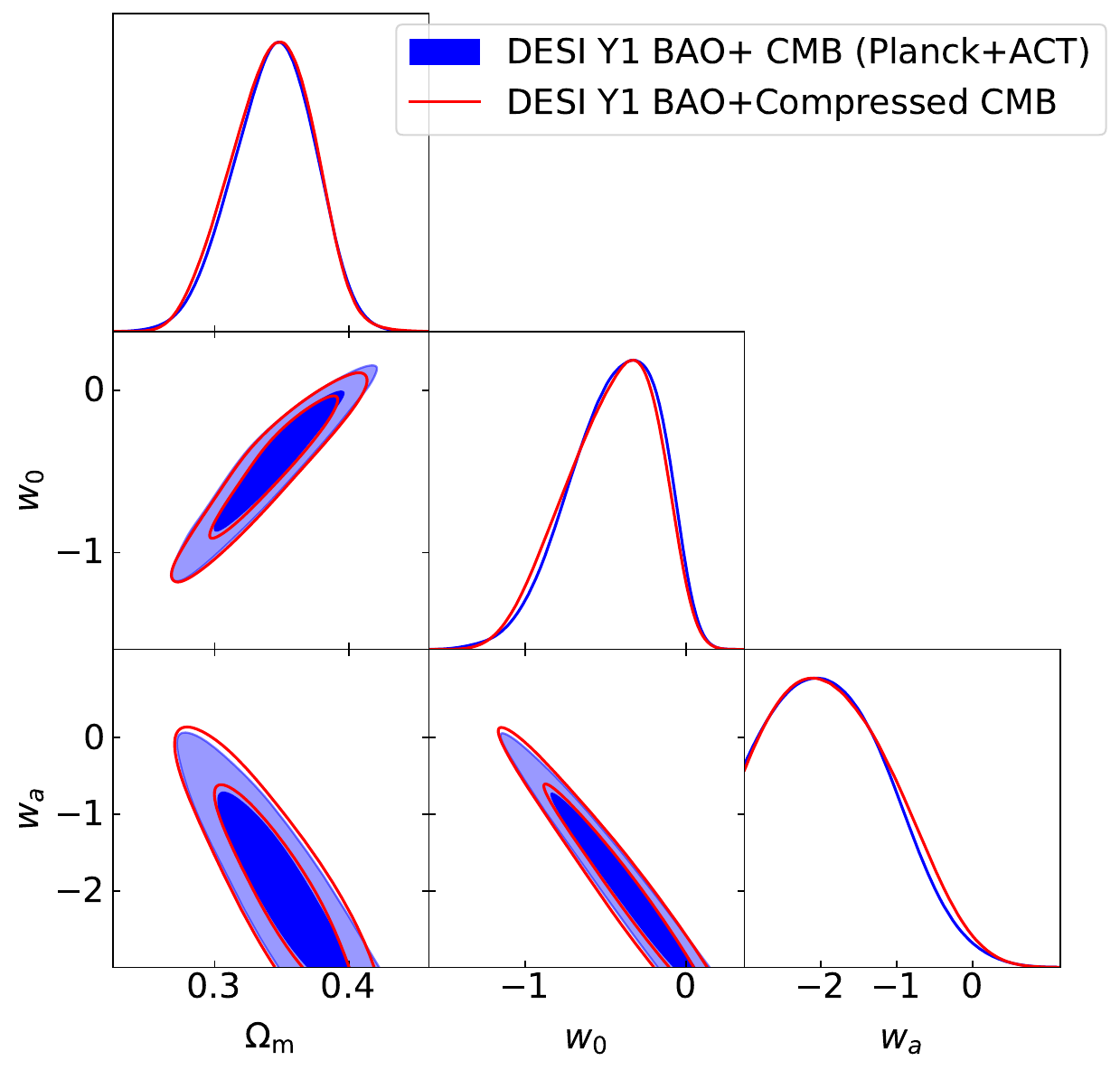}    
    \caption{Comparison of the combined CMB and DESI Y1 BAO constraints in the \lcdm\ model (left panel) and the \wowacdm \ model (right panel) to those obtained when the CMB likelihood is replaced with the compressed datavector and its covariance. The blue lines and filled contours show the constraints when we use the full \Planck\ + ACT chains for the CMB, while the red lines and open contours show the results with our compressed CMB datavector. }
    \label{fig:bao_cmb_constraints}
\end{figure*}

To test this conjectured increased robustness with data of increased constraining strength, we combine the same CMB data (\Planck\ (2018) + ACT (DR6)) with the BAO data from the first year of Dark Energy Spectroscopic Instrument (DESI Y1 BAO)\footnote{{We have explicitly checked that the performance of the compression does not depend on whether we use the DESI BAO data from Data Release 1 or Data Release 2; while we updated the results in the body of this paper with DR2 data \cite{DESI:2024mwx}, we show in these appendices the validation with DR1 which we carried out in an earlier version of this paper.}}. The left panel of Fig.~\ref{fig:bao_cmb_constraints} shows the expected good agreement between the exact and approximate treatment in \lcdm. The right panel, in turn, shows encouraging results from a much less trivial comparison in the \wowacdm\ model (on which, recall, the compression was not explicitly trained). We see an excellent agreement between the combined DESI Y1 BAO + CMB results, down to tracing subtle non-Gaussianities in the 2D parameter posteriors. 

The results just shown indicate that our compressed datavector faithfully represents the CMB information in cases when non-standard models affect the expansion history, as e.g.\ in the \wowacdm\ model. 

\begin{figure*}[t]
    \centering
    \includegraphics[width=0.49\linewidth]{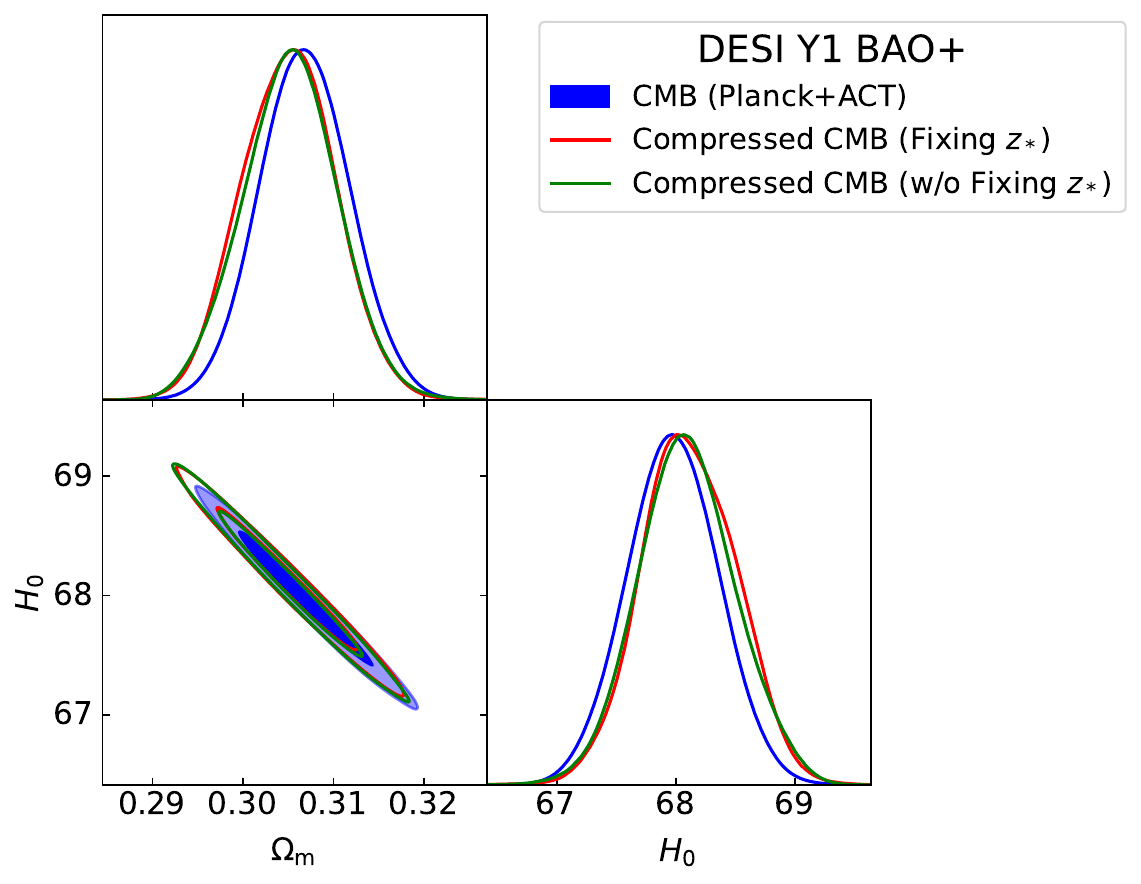}
    \includegraphics[width=0.49\linewidth]{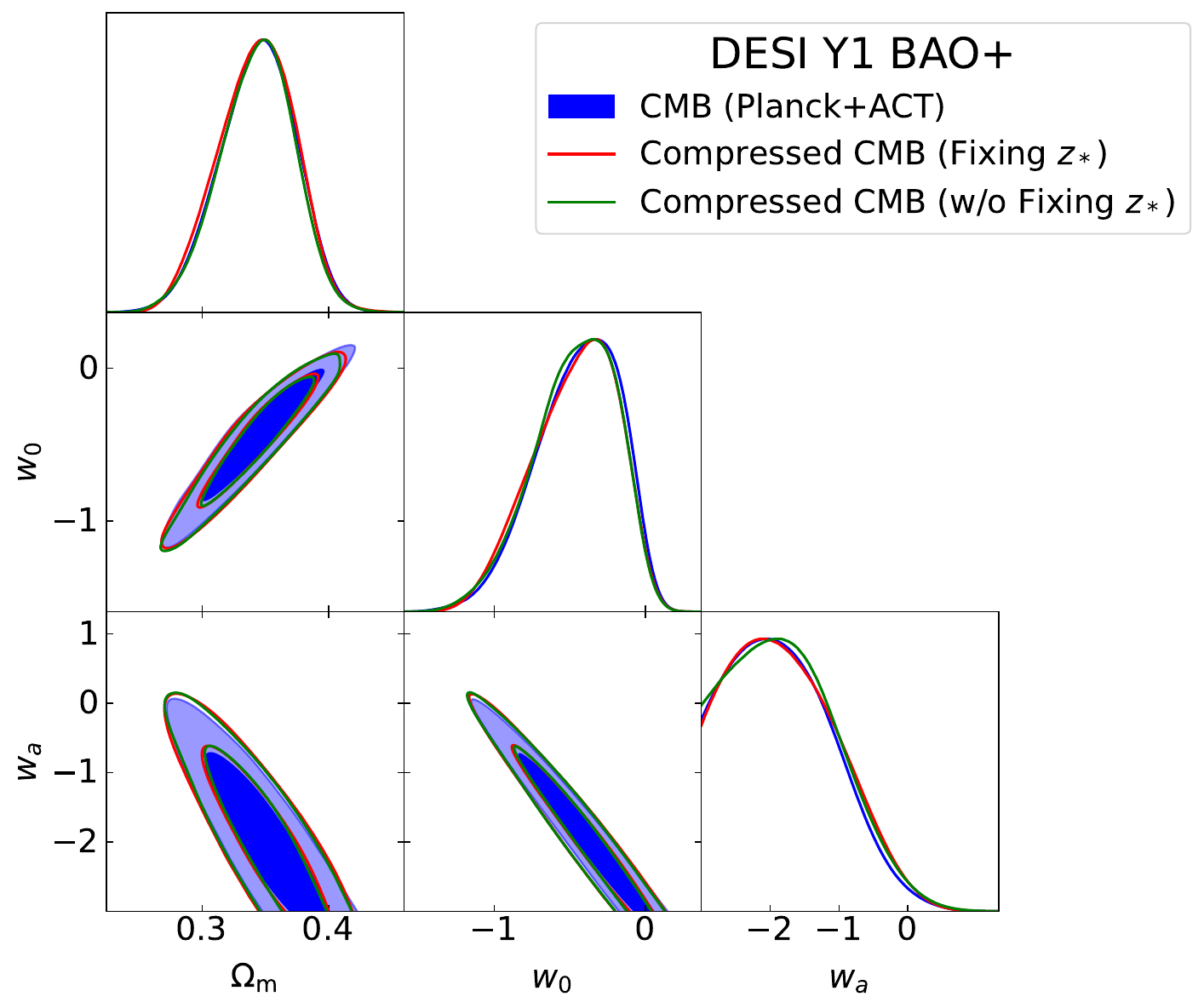}    
    \caption{Comparison of various CMB compression schemes combined with DESI Y1 BAO+ constraints in the $\Lambda$CDM model (left) and the $w_0w_a$CDM model (right). The blue filled contours and lines correspond to constraints using the full \textit{Planck} + ACT CMB chains. The red contours show the results using our compressed CMB data vector with $z_*$ fixed, while the green contours use the compressed data vector without fixing $z_*$.}
    \label{fig:fixing_zstar}
\end{figure*}

When computing the compressed CMB datavector and covariance, we fixed the redshift $z_*$ (photon decoupling surface) to $1090$ to calculate the means and covariances using the \texttt{CAMB} \cite{Lewis:1999bs} software. 
It is natural to question the validity of this assumption, as $z_*$ depends on the expansion history, and hence on the dark-energy model \cite{Hu_1996}. To address this issue, we recomputed the means and covariances using \texttt{CAMB} without fixing $z_*$. The derived quantities, \texttt{rstar} and \texttt{DAstar}, were calculated directly from \texttt{CAMB} and used to compute $R$ and $\ell_a$. Figure~\ref{fig:fixing_zstar} compares the constraints obtained using the two approaches for both the \lcdm\ and \wowacdm\ cases. Some references \cite{Gao:2024ily,Liu:2024fjy,Lazkoz:2023oqc} have suggested using the approximate expression for $z_*$ from Hu \& Sugiyama \cite{Hu_1996} for the compressed CMB likelihood. However, these expressions cannot be applied to exotic models that affect the thermal history of the universe. In such cases, our compression proves particularly useful, as it only requires computing the background quantities at $z_*=1090$ without any additional thermal history calculations to correct $z_*$.

\section{Constraints on the alphas}\label{app:alphas}

\begin{figure*}[t]
    \centering
    \includegraphics[width=0.9\linewidth]{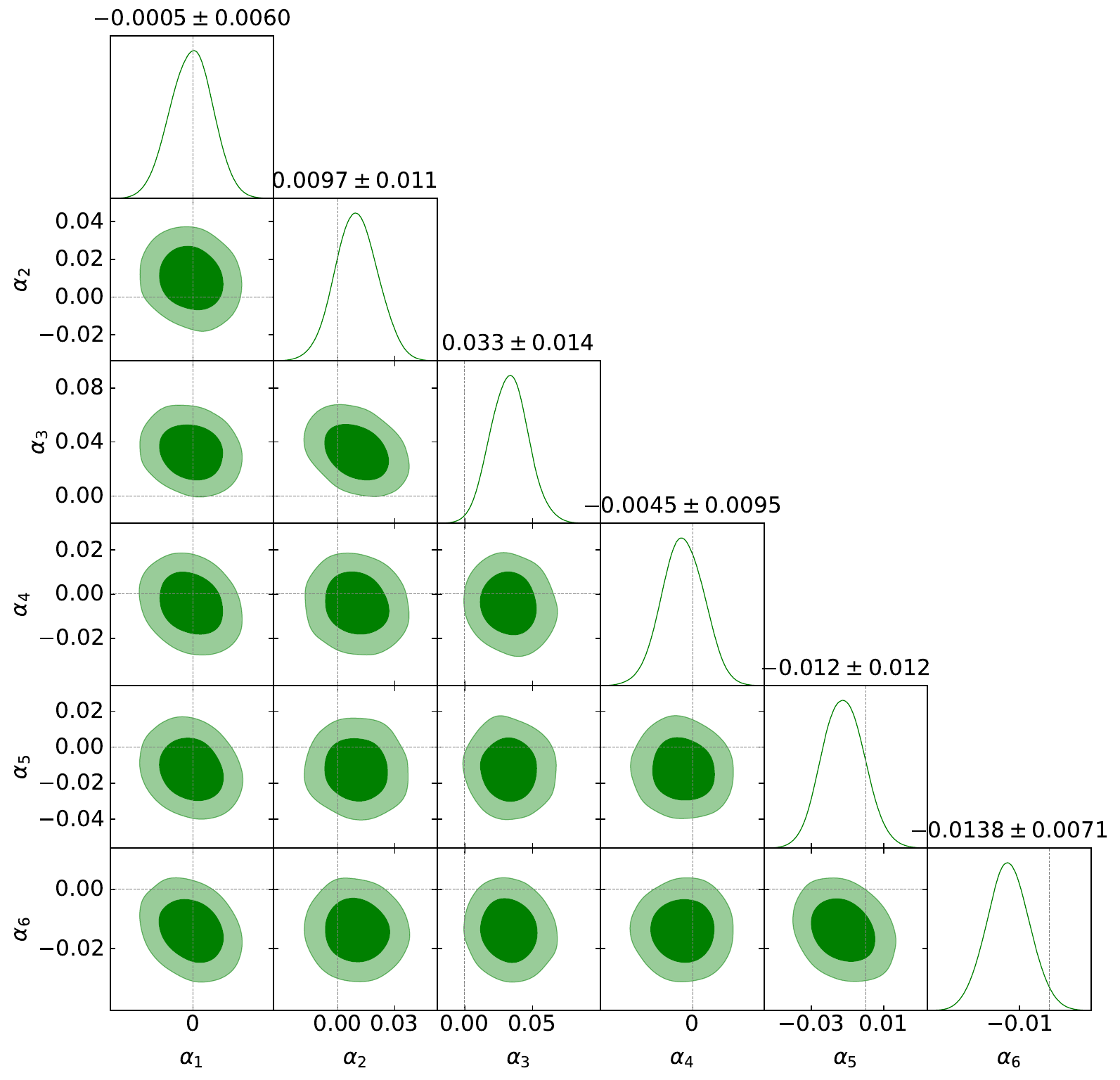}
    \caption{Constraints on the six parameters that describe perturbations in the Hubble rate (see Eq.~\ref{eq:Hz}), marginalized over the three remaining parameters ($\Obhtwo, \Ocdmhtwo$ and $\HnodLCDM$). The contours represent the 68.3\% and 95.4\% credible regions, and the numbers above each diagonal panel shows the projected mean and error in the corresponding alpha. The faint dashed lines are centered at fiducial values in the \lcdm\ model, which is zero for each of the alphas.}
    \label{fig:alphas}
\end{figure*}

Here we provide more detailed information about the constraints on the alpha parameters defined in our modified-H model (see Eq.~(\ref{eq:Hz})). Figure \ref{fig:alphas} shows the constraints on the six alpha parameters,
marginalized over the three remaining parameters ($\Obhtwo, \Ocdmhtwo$ and $\HnodLCDM$), assuming DESI+CMB+DESY5 data, while Table \ref{tab:constraints} shows the numerical constraints on all parameters. We see that each of the six alphas is constrained reasonably well; the errors {range from} $0.01$ to $0.04$, and this is how well the fractional expansion rate is constrained in the respective redshift bins. Note also that we iterated a little in selecting our flat priors on the alphas in order to ensure that none of the constraints hits the prior boundaries. Table \ref{tab:constraints} shows the numerical constraints on all eight parameters of our model, also assuming DESI+CMB+DESY5 data. [For clarity we do not show the constraints with the Union3 or PantheonPlus SNIa datasets.]

{We also comment on the calculation of $\Delta\chi^2_{\rm MAP}$ between the maximum a posteriori (MAP) modified-H and \lcdm\ model. This calculation is complicated by the fact that the MAP value for the modified-H model is challenging to locate (standard packages such as \texttt{iminuit} fail to find the minimum). We have instead computed these $\Delta\chi_{\rm MAP}^2$ values from the chains, and some further tests indicate that these numbers are reasonably accurate. In addition, we found that the compressed CMB results give slightly different values of $\Delta\chi_{\rm MAP}^2$ even in cases (e.g.\ \wowa\ model fits) where the minimization can be carried out. To give an example for the goodness of fit, a combination of DESI Y1 BAO, compressed CMB, and Union3 data (which is also compressed), altogether contain 37 measurements. For our fits with eight free parameters, we expect $\chi_{\rm MAP}^2$ of $29\pm \sqrt{2\times 29}\simeq 29\pm 8$, and we observe the best-fit of DESI+CMB+Union3 data in the modified-H model to be $\chi^2_{\rm MAP}\simeq 31$. }

\setlength{\tabcolsep}{1pt} 
\renewcommand{\arraystretch}{1.2} 
\begin{table}[t]
\centering
\caption{\label{tab:constraints} Constraints on the fundamental parameters of our modified-H model in our fiducial analysis of DESI+CMB+DESY5. 
}
\begin{ruledtabular}
\begin{tabular}{lc}
Parameter & Constraint\\  \hline  
$\HnodLCDM$\,\,\,\mbox{(value adopted from \lcdm\ fit)} & $68.24$ (Fixed)\\
$\Obhtwo$ & $0.02240\pm 0.00014 $\\
$\Ocdmhtwo$ & $0.1198\pm 0.0011 $\\
$\alpha_1$ & $-0.0005\pm 0.0060  $\\
$\alpha_2$ & $0.0097\pm 0.011     $\\
$\alpha_3$ & $0.033\pm 0.014         $\\
$\alpha_4$ & $-0.0045\pm 0.0095   $\\
$\alpha_5$ & $-0.012\pm 0.012    $\\
$\alpha_6$ & $-0.0138\pm 0.0071   $
\end{tabular}
\end{ruledtabular}
\end{table}

\section{Neutrino Density}
\label{app:mnu}

We have accounted for the neutrino density using the formalism developed in WMAP 7-year analysis \cite{Komatsu_2011}
\begin{equation}
    \Omega_\nu (a) = 0.2271\Omega_\gamma (a) N_\mathrm{eff}\left(\frac{1}{3}\sum_{i=1}^3f(m_{\nu,i} a/T_{\nu,0})\right)\label{eq:nu_komatsu}\,.
\end{equation}
Here $N_\mathrm{eff}$ is the effective number of neutrino species (fixed at a value 3.044 for this analysis), $a$ is the scale factor, $\Omega_\gamma$ is the photon density and $T_{\nu,0}$ is the neutrnio temperature at $z=0$ ($a=1$) given by

\begin{equation}
    T_{\nu,0} = \left(\frac{4}{11}\right)^{1/3}T_\mathrm{CMB} = 1.945 \mathrm{K}\,.
\end{equation}

The function $f$ represents the Fermi-Dirac integral given by
\begin{equation}
    f(y) = \frac{120}{7\pi^4}\int_0^\infty \frac{x^2\sqrt{x^2+y^2}}{e^x+1}.
\end{equation}
To make the code more efficient, we have used the fitting formula \cite{Komatsu_2011}
\begin{equation}
    f(y) \approx (1+(Ay)^p)^{1/p}
\end{equation}
where $A \approx 0.3173$ and $p=1.83$.

For the case of two massless neutrinos and one massive neutrino of mass $m_\nu=0.06$ \si{\electronvolt}, the term in the paranthesis in Eq.~(\ref{eq:nu_komatsu}) simplifies to
\begin{equation}
    \frac{1}{3}(2+f(m_{\nu} a/T_{\nu,0}))\,.
\end{equation}
Combining Eq.~(\ref{eq:nu_komatsu}) with the photon density, we can write the total radiation density (including the contribution from massive neutrinos) as
\begin{equation}
    \Or (a) = \Omega_\gamma (a) + \Omega_\nu (a)
\end{equation}

\bibliographystyle{apsrev4-2}
\bibliography{references}
\end{document}